\documentstyle[12pt,aasms4]{article}

\lefthead{Bekki Kenji}
\righthead{Formation of metal-poor gaseous halo  in gas-rich galaxy mergers
}

\begin{document}
\title{Formation of metal-poor gaseous halo  in gas-rich galaxy mergers}

\author{Kenji Bekki} 
\affil{Astronomical Institute, Tohoku University, Sendai, 980-8578, Japan \\
email address: bekki@astroa.astr.tohoku.ac.jp}

\begin{abstract}

We  numerically investigate  
chemodynamical evolution of interstellar medium (ISM) 
in gas-rich disk-disk galaxy mergers
in order to explore the origin of fundamental chemical properties of  halo ISM 
observed in  elliptical galaxies.
Main results obtained in this chemodynamical study are the following three.

(1) Elliptical galaxies formed by gas-rich  mergers  
show steep negative
metallicity  gradients of ISM especially in the outer part of galaxies.  
The essential reason for this is that
chemical evolution of ISM in mergers proceeds in such an inhomogeneous
way that in the central part of mergers, metal-enrichment of ISM is more efficient
owing to radial inflow of metal-enriched ISM 
during dissipative galaxy merging, whereas in the outer part,
metal-enrichment is less efficient owing to  a larger amount of metal-enriched
ISM tidally
stripped away from mergers.
This result provides a clue to the origin of gaseous metallicity
gradients of elliptical halo recently revealed by $ASCA$.

(2) Because of inhomogeneous chemical evolution of ISM in mergers,
$some$  merger remnants show mean gaseous metallicity discernibly smaller
than mean stellar one. 
The degree of difference in mean stellar and gaseous metallicity
in a merger remnant depends  on chemical mixing length,
galactic mass, and the effectiveness of supernova feedback.

(3) Elliptical galaxies formed by multiple mergers 
are more likely to have metal-poor gaseous halo components
and steep  gaseous metallicity gradients
than those formed by pair mergers.
This is principally because a larger amount of less-metal
enriched ISM can be tidally stripped away more efficiently from galaxies 
in multiple mergers.

These three  results demonstrate that dynamical evolution of gas-rich  galaxy mergers,
in particular, tidal stripping of less metal-enriched ISM during galaxy merging, 
greatly determines chemical evolution of ISM of galaxy mergers. 
These results furthermore imply that recent $ASCA$  observational results concerning
mean and radial chemical properties of halo ISM in elliptical galaxies
can be understood in terms of chemodynamical evolution of gas-rich
galaxy mergers.

\end{abstract}


\section{Introduction}
Recent observational studies by $ASCA$ 
($Advanced$ $Satellite$ $for$ $Cosmology$ $and$ $Astrophysics$) 
have  revealed a number of fundamental chemical properties
of interstellar medium (ISM) of elliptical galaxies,
thus have provided valuable information on the formation and 
evolution of elliptical galaxies
(e.g., Awaki et al. 1994; Loewenstein et al. 1994;
Matsushita et al. 1994; Mushotzky et al. 1994; Matsumoto et al. 1997;
Matsushita et al. 1997). 
For example, Fe abundance of
hot gaseous X-ray halo
has been revealed to be
appreciably  smaller than that of the stellar component in
the host elliptical galaxy (Awaki et al. 1994;
Matsushita et al. 1994; Matsumoto et al. 1997;
but see Matsushita et al. 1997).
This smaller gaseous metallicity 
is 
considered to be largely inconsistent with the theoretical prediction
of the conventionally  used one-zone chemical evolution model 
(`` The iron abundance discrepancy problem'')
and thus has been extensively discussed in theoretical studies
(e.g., Renzini et al.  1993; 
Fujita, Fukumoto, \& Okoshi 1996; Arimoto et al. 1997).
Furthermore, the hot $X$-ray gaseous halo in elliptical galaxies
has been revealed to show strong negative metallicity gradients,
which suggests that some physical mechanisms such
as cooling flow, gaseous dissipation, galaxy merging,
and dilution from external metal-poor gas play
a vital role in the formation of the gaseous metallicity gradients
(Loewenstein et al. 1994; Mushotzky et al. 1994;
Matsushita et al. 1997).
Although these $ASCA$ observational results on mean and radial chemical properties
of ISM can provide some diagnostics for any theories of elliptical
galaxy formation,
only a few theoretical studies have addressed the origin of the above fundamental 
characteristics of chemical properties of ISM in elliptical
galaxies.

The purpose of this paper is to 
explore the origin of fundamental chemical properties
of ISM of  elliptical galaxies recently revealed by $ASCA$.
We adopt the assumption that elliptical galaxies are
formed by gas-rich disk-disk galaxy mergers and  thereby 
investigate whether or not the merger model of elliptical galaxy formation
can give a plausible explanation for the origin of recent observational
results of $ASCA$ on mean and radial chemical properties of ISM in 
elliptical galaxies.
We particularly investigate
how the dynamical mixing of chemical components during galaxy merging,
which has not been investigated at all in  previous studies,
affects mean and radial  chemical properties of ISM 
of merger remnants.
In the present study, the key physical process associated with the
origin of metal poor gaseous halo and radial gaseous metallicity gradients
in elliptical galaxies
is demonstrated to be tidal stripping
of less metal-enriched  ISM  during galaxy merging.
Based on the present numerical results,
we  point out the disadvantages of the commonly used one-zone models
in discussing the metallicity of ISM  of elliptical galaxies
and stress the importance
of dynamical processes of galaxy formation
in the chemical enrichment processes of ISM in galaxies.
The layout of this paper is as follows.
In \S 2, we summarize  numerical models used in the
present study.
In \S 3, we demonstrate how  a number of fundamental $chemical$
properties of ISM 
in merger remnants
are affected by purely $dynamical$ processes of galaxy merging.
In \S 4, we provide a number of implications on chemical properties
of ISM in elliptical galaxies.
The conclusions of the present study are given in \S 5.

\section{Model}
We mainly investigate a number of fundamental chemical properties
of ISM  in ellipticals
formed by  galaxy merger between two gas-rich spirals.
Fundamental chemical properties of ISM  in galaxy mergers are basically  
controlled by the star formation histories,
which are considered to strongly depend on the
time-evolution of dynamical and kinematical properties
such as local dynamical instability  in galaxies (e.g. Kennicutt 1989;
Larson 1992).
Hence we numerically solve the time-evolution of chemical
evolution of galaxy mergers, based on
the dynamical evolution of galaxies.
Since the details of the chemodynamical model is given in Bekki \& Shioya (1998),
we only briefly describe the 
model in the present study. 
We first describe a  numerical model for dynamical evolution of
galaxy mergers, including structure and kinematics of merger progenitor
disks and star formation law (\S 2.1),
and then give the method for analyzing the chemical enrichment process
during galaxy merging (\S 2.2).

\subsection{A model  of gas-rich galaxy mergers}

 We construct  models of galaxy mergers between gas-rich 
 disk galaxies with equal mass by using Fall-Efstathiou model (1980).
 The total mass and the size of a progenitor disk are $M_{\rm d}$
 and $R_{\rm d}$, respectively. 
 From now on, all the mass and length are measured in units of
  $M_{\rm d}$ and  $R_{\rm d}$, respectively, unless specified. 
  Velocity and time are 
  measured in units of $v$ = $ (GM_{\rm d}/R_{\rm d})^{1/2}$ and
  $t_{\rm dyn}$ = $(R_{\rm d}^{3}/GM_{\rm d})^{1/2}$, respectively,
  where $G$ is the gravitational constant and assumed to be 1.0
  in the present study.
  Dimensional values for these  units in each model are given later.
  In the present model, the rotation curve becomes nearly flat
  at  0.35  $R_{\rm d}$  with the maximum rotational velocity $v_{\rm m}$ = 1.8 in
  our units.
  The corresponding total mass $M_{\rm t}$ and halo mass $M_{\rm h}$
  are 3.8 and 2.8 in our units, respectively.
  The radial ($R$) and vertical ($Z$) density profile 
  of a  disk are  assumed to be
  proportional to $\exp (R/R_{0}) $ with scale length $R_{0}$ = 0.2
  and to  ${\rm sech}^2 (Z/Z_{0})$ with scale length $Z_{0}$ = 0.04
  in our units,
  respectively.
  In addition to the rotational velocity made by the gravitational
  field of disk and halo component, the initial radial and azimuthal velocity
  dispersion are given to disk component according
  to the epicyclic theory with Toomre's parameter (\cite{bt87}) $Q$ = 1.0.
  The vertical velocity dispersion at given radius 
  are set to be 0.5 times as large as
  the radial velocity dispersion at that point, 
  as is consistent  with 
  the observed trend  of the Milky Way (e.g., Wielen 1977).
  The collisional and dissipative nature 
  of the interstellar medium (ISM) is  modeled by the sticky particle method
  (\cite{sch81}). 
  The size  of the clouds is set to be 5.0 $\times$ $10^{-3}$ in our units 
  in the present simulations. 
  The radial and tangential restitution coefficient for cloud-cloud
  collisions are
  set to be 0.5 and
  0.0, respectively.
  We assume that   the fraction of gas mass in
  a disk is set to be 1.0 initially, in order to 
  mimic higher redshift 
  galaxy mergers which probably have a considerably larger amount of ISM. 
  The reason of this  assumption is that recent observational studies such as  
  the tightness of the color-magnitude relation 
  in the  cluster of galaxies (e.g., Bower
  et al. 1992, Ellis et al. 1997) suggest the relatively higher redshift of typical
  elliptical galaxy formation.

    Star formation
     is modeled by converting  the collisional
    gas particles
    into  collisionless new stellar particles according to the algorithm
    of star formation  described below.
    We adopt the Schmidt law (Schmidt 1959)
    with exponent $\gamma$ = 2.0 (1.0  $ < $  $\gamma$
      $ < $ 2.0, \cite{ken89}) as the controlling
    parameter of the rate of star formation.
    The amount of gas 
    consumed by star formation for each gas particle
    in each time step, 
    $\dot{M_{\rm g}}$, 
is given as:
    \begin{equation}
      \dot{M_{\rm g}} \propto  
 {(\rho_{\rm g}/{\rho_{0}})}^{\gamma - 1.0}
    \end{equation}
    where $\rho_{\rm g}$ and $\rho_{0}$
    are the gas density around each gas particle and
    the mean gas density at 0.48 radius  of 
    an initial disk, respectively.
Thermal and dynamical (kinematic) heating of ISM  driven by
supernovae are found to affect greatly  the formation and
evolution of galaxies
(e.g., Katz 1992; Navarro \& White 1993). 
We here consider  the dynamical  feedback effects of Type Ia 
(SNIa) and Type II supernovae (SNII) and neglect the feedback effects
of thermal heating on galaxy evolution.
The reason for this neglection
is firstly that our gaseous model does not solve thermal evolution
of ISM, and secondly that  such thermal effects is found to be less
important than dynamical feedback effects owing to efficient cooling
of ISM (e.g., Katz 1992; Navarro \& White 1993).
10 percent of total energy ejected from SNIa and SNII is  assumed to  
be returned into ISM for giving  velocity perturbation of  ISM in the present study.
More details of the supernovae feedback effects are given in Bekki \& Shioya (1998).

 We investigates two different types of  merger models,
both of which  are considered to be promising candidates
of elliptical galaxy formation (Barnes \& Hernquist 1992):
Pair mergers between two disks
and multiple mergers between  five disks. 
The initial orbit conditions of two merger progenitor disks for the pair
merger model are as follows.
The  orbit of the two disks in a pair merger is set to be
initially in the $xy$ plane.
The initial distance between
the center of mass of the two disks,
the pericenter
distance,
and the eccentricity of merger orbit are 4.0, 1.0, and 1.0, respectively.
The spin of each galaxy in a  pair merger
is specified by two angle $\theta_{i}$ and
$\phi_{i}$, where suffix  $i$ is used to identify each galaxy.
$\theta_{i}$ is the angle between the $z$ axis and the vector of
the angular momentum of a disk.
$\phi_{i}$ is the azimuthal angle measured from $x$ axis to
the projection of the angular momentum vector of a disk on
to $xy$ plane.
$\theta_{1}$, $\theta_{2}$, $\phi_{1}$, and $\phi_{2}$
are set to be 30.0, 120.0, 90.0, and 0.0, respectively.
In the simulations of multiple mergers, the initial position of
each progenitor disk is set to be distributed 
randomly within a sphere with radius 6.0 in
our units, and the initial velocity dispersion of each disk (that is,
the random motion of each galaxy in the sphere) is set to be
distributed  in such a
way that the ratio of the total kinematical 
energy to the total potential energy in the system is  0.25.  
The time-scale in which the progenitor disks merge completely and reach  the
dynamical equilibrium of an elliptical 
galaxy is less than 20.0 in our units for the two 
models.

   All the calculations related to 
the above dynamical evolution  including the dissipative
dynamics, star formation, and gravitational interaction between collisionless
and collisional component 
 have been carried out on the GRAPE board
   (\cite{sug90})
   at Astronomical Institute of Tohoku University.
  The  number of particles of halo and the gaseous component
  are 10000 and 20000 for the  pair merger model, and
  25000 and 25000 for the multiple merger one.
   The parameter of gravitational softening is set to be fixed at 0.04  
   in all the simulations. The time integration of the equation of motion
   is performed by using 2-order
   leap-flog method. 
   Energy and angular momentum  are conserved
within 1 percent accuracy in a test collisionless merger simulation.
Most of the  calculations are set to be stopped at T = 20.0 in our units
(corresponding to a few Gyr)
unless specified.

\subsection{Chemical enrichment}

 Chemical enrichment through star formation during galaxy merging
is considered  to proceed  locally and inhomogeneously in the present
chemodynamical model.
We investigate time-evolution
of  five species of chemical components,
H, He, O, Mg, and Fe as well as conventionally used mean
metallicity Z. 
The mean metallicity of Z
for each $i$th stellar particle is represented by $Z_{i}$. 
Total mass of each $j$th  ($j$=1,2,3,4,and 5) chemical component (H, He, O, Mg, and Fe) 
ejected from  each $i$th stellar particle through SNIa and SNII at the time $t$ 
is given as, 
\begin{equation}
\Delta z_{i,j} (t) = m_{\rm s, \it i}Y_{i,j}(t-t_{i}), 
\end{equation}
where  $m_{\rm s, \it i}$ is the mass of the $i$th stellar particle,
$Y_{i,j}(t-t_{i})$ is the mass of each $j$th chemical component  per unit mass
at the time $t$,
and $t_{i}$ represents the time when the $i$th stellar particle is
born from a gas particle.
The $\Delta z_{i,j} (t)$ is given to
neighbor gas particles locating within
$R_{\rm chem}$ from the position of the $i$th  stellar particle.
This  $R_{\rm chem}$ is referred to as 
chemical mixing length in the present paper,
and represents the region within which the neighbor
gas particles are polluted by metals ejected from  stellar particles.
The value of  $R_{\rm chem}$  relative to the typical size of
a galaxy could be different between galaxies, accordingly the
value of  $R_{\rm chem}$  is considered to be free parameter in
the present study.
The value of  $R_{\rm chem}$ examined the most extensively 
in the present study is 
0.1, which corresponds to the half of the scale-length of initial disks.
Initial gaseous metallicity for each chemical components
is set to be 0.1 solar, which is exactly the same as that
of infall gas adopted in the classical chemical evolution models of disk galaxies
with gaseous infall.
Although we investigate the above six species of chemical components
(H, He, O, Mg, Fe, and Z), we only present the results of  Fe and Z 
in the present study.

Since we now consider the time-delay between the epoch of star formation
and that of supernovae explosions, the mass
of each $j$th chemical component ejected from each
$i$th stellar particle through SNIa and SNII,
which are basically  determined  by $Y_{i,j}(t-t_{i})$,
is also strongly time-dependent.
We  estimate the mass-dependent life-time 
of stars that becomes SNIa or SNII 
by using mass-age relation given by Bressan et al (1993).
The fraction of close binary stars 
of SNIa relative to SNII 
(represented by  $A$ parameter
in Matteucci \& Tornamb\`{e} 1987)
is assumed to be 0.1.
The $Y_{i,j}(t-t_{i})$ furthermore depends on stellar yields, IMF profiles,
upper cutoff mass $M_{\rm up}$, and lower one $M_{\rm low}$.
In the present study, 
we adopt the Salpeter  (IMF),
$\phi(m) \propto m^{-1.35}$,
with $M_{\rm up}=120 M_{\odot}$ and
$M_{\rm low}=0.6 M_{\odot}$.
The reason for this larger $M_{\rm low}$ (=$0.6 M_{\odot}$)
is essentially because we do not have stellar yield tables
for stars with  masses less than 0.6.
To calculate the ejected mass of gas and metals in $Y_{i,j}(t-t_{i})$,
we use stellar yields derived by  Woosley \& Weaver (1995) for SNII,
Nomoto, Thielemann, \& Yokoi (1984) for  SNIa,
and Bressan et al. (1993)
and  Magris \& Bruzual (1993) for low and intermediate mass stars.
More details of the time-dependence of $Y_{i,j}(t-t_{i})$ for a given
IMF, $M_{\rm up}$, and $M_{\rm low}$ are  given in Bekki \& Shioya (1998).

\placetable{tbl-1}
\placefigure{fig-1}
\placefigure{fig-2}

\subsection{Model parameters}
 By using the above chemodynamical model of gas-rich galaxy mergers, 
 we mainly investigate
 the following two points concerning chemical evolution of ISM:
 (1) The radial chemical properties of ISM (Z and Fe) in merger remnants,
 and (2) The degree of difference in $mean$ $metallicity$ between
 stellar components and gaseous ones (ISM) in merger remnants.
 We describe numerical  results of the following five models (Model 1,2,3,4, and 5):
 Model 1 is a pair merger model with $M_{\rm d} = 10^{10} M_{\odot}$
 and $R_{\rm chem}=0.1$,
 which shows a typical behavior of chemical evolution
 of ISM in the present merger model  thus is referred to as the standard model;
 Model 2 is a pair merger model with $M_{\rm d} = 10^{12} M_{\odot}$
 and $R_{\rm chem}=0.1$,
 which describes the results of more massive galaxy  mergers;
 Model 3 is a pair merger model with $M_{\rm d} = 10^{10} M_{\odot}$
 and $R_{\rm chem}=0.4$,
 in which chemical evolution of ISM is assumed to proceed more globally and homogeneously
 owing to the larger chemical mixing length;
 Model 4 is a pair merger model with $M_{\rm d} = 10^{10} M_{\odot}$
 and $R_{\rm chem}=0.1$,
 in which total amount of supernova energy that is returned into 
  ISM is five times larger than that adopted in Model 1 (a model
 with stronger supernova feedback);
 Model 5 is a $multiple$  merger model with $M_{\rm d} = 10^{10} M_{\odot}$
 and $R_{\rm chem}=0.1$,
 in which  five disks are assumed to merge with each
 other to form an  elliptical galaxy.
The dimensional value  of  $T_{\rm dyn}$  (units of the present study)
and that of  $R_{\rm d}$ 
are $9.0 \times 10^{7}$ yr and $7.4$ kpc, respectively, 
for models with $M_{\rm d} = 10^{10} M_{\odot}$,
and  $2.9 \times 10^{8}$ yr and $74.1$ kpc, respectively,
for models with $M_{\rm d} = 10^{12} M_{\odot}$.
The parameter values for these five models
are summarized  in Table 1.

\section{Results}

\subsection{Inhomogeneous chemical evolution of ISM in mergers}

In the present chemodynamical model of gas-rich galaxy mergers,
metals produced in star-forming regions of mergers are assumed to be 
mixed only $locally$ into the surrounding ISM.
It is just this $local$ mixing of chemical components that plays a vital
role in determining not only $radial$ but also $mean$ chemical properties
of ISM in merger remnants.
Figure 1 describes radial properties of mean gaseous metallicity,
$<Z>$ and $<Fe>$, in a merger remnant at 
the time $T=16.0 T_{\rm dyn}$ (corresponding to $\sim$ 1.4 Gyr)
for Model 1 (the standard model).
It is clear from Figure 1 that the merger remnant has steep 
negative metallicity gradient of ISM both for $<Z>$ and $<Fe>$.
Furthermore gaseous metallicity in the outer part of the merger remnant
is found to be considerably smaller ($\sim$ 0.5 solar),
which implies  that the formation of metal-poor gaseous halo
surrounding the outer part of galaxies is an inevitable physical process for ellipticals
formed by gas-rich galaxy mergers.
These results suggest that $dynamical$ evolution of gas-rich galaxy merging
greatly affects $chemical$ evolution of ISM and thus controls radial properties
of $<Z>$ and $<Fe>$.
The details of the  formation process of gaseous metallicity gradients 
are given  as 
follows.
In the present chemodynamical model of gas-rich galaxy mergers,
metals produced and ejected by SNIa and SNII in star-formating gaseous 
regions of a merger 
can be mixed only locally into the surrounding ISM (`Inhomogeneous
chemical mixing').
Accordingly, the metals, which are mostly produced in the central region
of the  merger,  can be mixed preferentially into 
ISM  in the central region where further efficient star formation
is expected, thus can not be mixed so efficiently into
the outer region of the merger.
Consequently, the ISM  of the outer part of the
merger remains less metal-enriched.
Such less metal-enriched ISM in the outer part of the merger 
is   then  effectively stripped away from the system during tidal interaction
of galaxy merging and finally  transferred to the more outer region
where  metals produced in the central part of the  merger
are harder to be mixed into.  
As a natural result of this, the mean metallicity 
of the ISM remaining in the outer part of the remnant
becomes   considerably smaller.
On the other hand, the ISM initially located in the central part of
the  merger 
can be more metal-enriched 
owing to quite efficient star formation there. 
The formation of gaseous metallicity gradients 
accordingly reflects  the fact
that the  metals produced by star formation  
are more efficiently trapped by the ISM in the central part of mergers than by
that  in the outer part.
Thus, the origin of negative metallicity gradients in the gaseous haloes of  merger remnants
is due principally to the inhomogeneous chemical mixing in gas-rich mergers.

The above inhomogeneous mixing of metals furthermore plays a decisive role
in determining $mean$ chemical properties of ISM in merger remnants.
Figure 2 describes the dependence of mean stellar metallicity
($<Z_{\ast}>$ and $<Fe_{\ast}>$) on mean gaseous one ($<Z>$ and $<Fe>$)
at  the time $T=8.0, 12.0, 16.0 T_{\rm dyn}$
in a galaxy merger for Model 1. 
Mean gaseous and stellar metallicity are measured 
for the region $R \leq R_{\rm eff}$ (where $R_{\rm eff}$ is the effective
radius of the remnant), 
$R \leq 2.5R_{\rm eff}$,
$R \leq 10.0R_{\rm eff}$,
and the whole region, in the merger remnant. 
As is shown in  Figure 2,
$<Z>$ is larger than $<Z_{\ast}>$ for $R \leq R_{\rm eff}$
whereas $<Z>$ is discernibly smaller  than $<Z_{\ast}>$ for the whole region.
This result implies that metals produced by star formation can be more
homogeneously  mixed into ISM in the inner regions of galaxy mergers.
$<Fe>$ is found to be larger than  $<Fe_{\ast}>$ for the above four regions. 
Furthermore the difference in mean stellar and gaseous metallicity
depends on the region for which mean metallicity is measured 
in such a way that the difference is larger in the more central  region.
What is the most significant among these results is that 
mean gaseous metallicity  $<Z>$ can be smaller than
mean stellar one  $<Z_{\ast}>$ in merger remnants:
Classical one-zone chemical evolution models never fails to  
predict that mean gaseous metallicity is always larger 
than (or equal to) mean stellar one (e.g., Arimoto \& Yoshii 1987).
Thus the present chemodynamical model suggests
that dynamical evolution of galaxy mergers,
in particular, tidal stripping of less metal-enriched
ISM during merging, can determine even mean chemical properties
of stellar and gaseous components  of merger remnants.

\placefigure{fig-3}
\placefigure{fig-4}

The important roles of inhomogeneous chemical mixing of metals
in determining mean and radial chemical properties of ISM 
can be more discernibly observed in a multiple galaxy merger than
in pair ones.
Figure 3 describes radial properties of mean gaseous metallicity,
$<Z>$ and $<Fe>$, in a  remnant of a multiple galaxy merger at 
the time $T=16.0 T_{\rm dyn}$ (corresponding to $\sim$ 1.4 Gyr)
for Model 5.
As is shown in Figure 3, the  remnant of the multiple galaxy merger  
has negative gradients of $<Z>$ and $<Fe>$ appreciably steeper
than those of the pair merger in Model 1 (Compare Figure 1 with  Figure 3.).
The reason for this steeper metallicity gradients is
that a larger amount of less metal-enriched ISM
can be tidally stripped away from galaxies owing to stronger
multiple galaxy interaction and finally
surrounds the outer part of merger remnants in multiple mergers.
Figure 4 describes the dependence of mean stellar metallicity
($<Z_{\ast}>$ and $<Fe_{\ast}>$) on mean gaseous one ($<Z>$ and $<Fe>$)
at  the time $T=8.0, 12.0, 16.0 T_{\rm dyn}$
in a multiple galaxy merger model (Model 5). 
As a natural result of stronger tidal stripping of less metal-enriched
ISM in the multiple galaxy merger,
gaseous metallicity averaged out for the whole region of the merge remnant
in Model 5 is appreciably smaller than stellar one.
These results thus confirm the importance of dynamical evolution of galaxy mergers
in determining mean and radial chemical properties of ISM in merger remnants. 

\placefigure{fig-5}

\subsection{Parameter dependence}

Figure 5 describes the dependence of mean stellar metallicity
($<Z_{\ast}>$ and $<Fe_{\ast}>$) on mean gaseous one ($<Z>$ and $<Fe>$)
at  the time $T=16.0 T_{\rm dyn}$ 
in each  merger model (Model 1 $\sim$ 5). 
From Figure 5, we can clearly observe the following five points.
Firstly, irrespectively of model parameters,
mean stellar metallicity ($<Z_{\ast}>$ and $<Fe_{\ast}>$)
is smaller  than mean gaseous one ($<Z>$ and $<Fe>$) in  the central
part of merger remnants ($R \leq R_{\rm eff}$).
This result implies that for all merger models,
metals produced in  star-forming regions of mergers
can be more efficiently and homogeneously
mixed into the surrounding ISM for the central part of galaxy mergers.
Secondly, $<Fe>$ is larger than $<Fe_{\ast}>$ both for the region
$R \leq R_{\rm eff}$ and for the whole region of merger remnants.
Thirdly, mean gaseous metallicity $<Z>$ can be smaller
than mean stellar one $<Z_{\ast}>$ especially for more massive
galaxy mergers and multiple galaxy mergers.
This result suggests that  $some$ ellipticals 
with  metal-poor gaseous halo are more likely to be remnants
of multiple galaxy mergers between more massive disks. 
Fourthly, even  (mean) gaseous metallicity $<Z>$ averaged out for the whole
region of merger remnants is larger 
than the corresponding  (mean) stellar one $<Z_{\ast}>$ in the model with
larger chemical mixing length (Model 3).
This is essentially because  in Model 3, chemical mixing
of metals proceeds more homogeneously
during galaxy merging owing to larger chemical mixing length adopted in Model 3. 
Fifthly, although mean gaseous metallicity is considerably smaller
in Z and Fe in the model with stronger feedback of SNIa and SNII (Model 4),
the mean gaseous metallicity is still 
discernibly larger than mean stellar one in Model 4.
This result implies that effectiveness of supernova feedback
does not strengthen  the difference in mean stellar and gaseous metallicity
in merger remnants thus gives  negligible effects on
the formation of gaseous halo with the mean metallicity smaller than the stellar
one.
These five results imply that although certain physical conditions of 
gas-rich galaxy mergers are required,
mean gaseous metallicity of ISM   
can be discernibly smaller
than mean stellar one in $some$ merger remnants.

\section{Discussion and conclusions}

 There are a growing number of observational evidences which
 suggest strong radial negative gradients of hot X-ray gaseous halo
 in elliptical galaxies
 (Loewenstein et al. 1994; Mushotzky et al. 1994;
 Matsushita et al. 1997).
 Observational study of NGC 4636 (Matsushita et al. 1997) have revealed 
 a factor of $3\sim4$ difference of  ISM metallicity within  
 $\sim 7.1 R_{\rm eff}$. 
 Cooling flows,  stellar metallicity gradients actually existing in ellipticals,
 the long-term chemical evolution of ISM driven by stellar mass
 loss or SNIa, 
 and dilution of metal-enriched ISM by external metal-poor gas are considered
 to be likely explanations for the origin of gaseous metallicity gradients
 (e.g., Loewenstein et al. 1994; Mushotzky et al. 1994;
 Matsushita et al. 1997).
 These likely explanations are closely associated either with external metal-poor gas 
 or with the later chemical evolution of ellipticals. 
 The present study provides an alternative explanation for  the origin:
The ISM  metallicity gradients can be closely associated with
 intrinsic and dynamical  processes of dissipative
 galaxy merging at the epoch of elliptical formation.
Inhomogeneous and radial-dependent chemical evolution of
galaxies  is found to play decisive roles in the formation
of negative gaseous metallicity gradients of merger remnants.
The present numerical results thus imply that the present-day
metallicity gradients of hot ISM halo can contain a fossil record
of the past dynamical evolution of ISM of elliptical galaxies 
at the epoch of their formation.

Negative metallicity gradients 
derived in
the present study are only true for ellipticals with a few Gyr age, 
primarily because 
we only solved
a few Gyr evolution of chemical properties of ISM 
but did not solve the later long-term
(corresponding to the Hubble time)
evolution. 
Thus, we should investigate the following two points in our future
studies in order to
confirm whether or not the proposed explanation for the origin
of observed metallicity gradients is actually viable
for the present-day ellipticals with their ages of  $\sim$ 10 Gyr  
in a more quantitative sense.
The first is 
long-term chemical evolution of ISM surrounding merger remnants, 
which can be  greatly affected by 
the later and continuous metal-enrichment due to SNIa 
and stellar mass loss of long-lived stars.
As has been demonstrated by several previous studies on 
long-term hydrodynamical evolution of hot ISM in ellipticals
(e.g., David, Forman, \& Jones 1991),
the effectiveness of thermal heating by SNIa 
(and partly by long-lived stars)
determines time-evolution radial flow patters  of 
hot ISM (e.g., either outflow
due to effective  thermal heating of SNIa 
or inflow due to efficient cooling).
Such later gaseous inflow or outflow,
which can transfer metal-enriched ISM,
can change radial metallicity distribution that is initially
formed by gas-rich galaxy merging.
Accordingly, chemodynamical effects of the later stellar mass-loss
and supernovae on the radial metallicity distribution
should be explored more in detail in our future studies.
The second is the long-term dynamical evolution
of less metal-enriched ISM that is tidally stripped and surrounds merger remnants.
The present study predicts that 
the tidally stripped metal-enriched ISM surrounds the considerably
outer part of
merger remnants, where external tidal field resulting from 
neighbor galaxies and  large-scale gravitational potential of cluster
or group of galaxies can affect dynamical evolution of the stripped ISM.
Later dynamical effects of external tidal force can change drastically
initial radial gradients of ISM in merger remnants,
thus we should also consider  these in our future studies.
Since radial  metallicity gradients of ISM in ellipticals
contain valuable information not only on chemical evolution
of ISM but also on dynamical evolution of galaxies as a whole,
more extensive studies are necessary for 
the better understanding of the origin of the gradients.

Furthermore,  the difference in mean metallicity between stellar and gaseous
components derived in the present merger model can be compared with 
recent observational results by $ASCA$ 
(Awaki et al. 1994;
Matsushita et al. 1994; Matsumoto et al. 1997)
which reveals that
the abundance of
hot gaseous X-ray halo (Fe)
is appreciably  smaller than that of the stellar component in
the host elliptical galaxy.
These observational results seem to be consistent with
results derived  in some merger models of the present study,
which imply that metal-poor gaseous components in some elliptical haloes
can be formed by gas-rich mergers.
The most recent result of Matsushita et al. (1997), however,
shows that there is not so large difference between
mean stellar metallicity  (0.74 solar) and mean 
gaseous one ($\sim$ 1.0 solar) within $\sim 4.0 R_{\rm eff}$
of NGC 4636 
and furthermore that the gaseous metallicity
in the outer part of the halo (for the region $4.7 \leq R/R_{\rm eff} \leq 7.1$) is 
still smaller ($\sim 0.37$  solar) than mean stellar metallicity of NGC 4636.
Although  metallicity averaged out for the whole region of the gaseous halo 
in NGC 4636 has not been clarified yet, 
this  new observational result provides the following implication on 
inhomogeneous chemical mixing derived in the present study.
If the gaseous metallicity averaged out
for the whole region of the gaseous halo 
is really larger than stellar one in NGC 4636,
the present numerical result that mean gaseous metallicity
can be smaller than mean stellar one in some merger remnants
is not consistent  with the observational
result.
In this case, we should consider either that inhomogeneous 
chemical mixing is not so efficient in real 
galaxy mergers as the present study predicts,
or that the later  long-term metal-enrichment
of ISM resulting from metal-ejection from long-lived stars and supernovae (SNIa)
can greatly affect the chemical evolution of ISM after galaxy merging
and thus change the difference of  mean stellar and gaseous metallicity
in merger remnants.
Alternatively, if the gaseous metallicity averaged out
for the whole region of the gaseous halo
is really  smaller  than stellar one,
the present study can provide  a clue to the origin
of such  smaller gaseous metallicity of elliptical haloes:
Origin of metal-poor gaseous halo in ellipticals
can be closely associated with the past dissipative
galaxy merging processes.
Since  total number of sample X-ray gaseous halo with the mean metallicity 
estimated precisely is still small,
it is safe for us to say, at least now, that
future more extensive observational studies 
and more elaborated theoretical models will assess  the validity
of inhomogeneous chemical mixing derived in the present study. 

Main results obtained in the present  chemodynamical study are the following three.

(1) Elliptical galaxies formed by gas-rich  mergers  
show steep negative
metallicity  gradients of ISM especially in the outer part of galaxies.  
The essential reason for this is that
chemical evolution of ISM in mergers proceeds in such an inhomogeneous
way that in the central part of mergers, metal-enrichment of ISM is more efficient
owing to radial inflow of metal-enriched ISM 
during dissipative galaxy merging, whereas in the outer part,
metal-enrichment is less efficient owing to  a larger amount of metal-enriched
ISM tidally
stripped away from mergers.
This result provides a clue to the origin of gaseous metallicity
gradients of elliptical halo recently revealed by $ASCA$.

(2) Because of inhomogeneous chemical evolution of ISM in mergers,
$some$  merger remnants show mean gaseous metallicity discernibly  smaller
than mean stellar one. 
The degree of difference in mean stellar and gaseous metallicity
in a merger remnant depends  on chemical mixing length,
galactic mass, and the effectiveness of supernova feedback.

(3) Elliptical galaxies formed by multiple mergers 
are more likely to have metal-poor gaseous halo components
and steep  gaseous metallicity gradients
than those formed by pair mergers.
This is principally because a larger amount of less-metal
enriched ISM can be tidally stripped away more efficiently from galaxies 
in multiple mergers.

These three  results demonstrate that dynamical evolution of galaxy mergers
can greatly affect chemical evolution of ISM of galaxies, 
which cannot be attained untill
both dynamical and chemical evolution of galaxies are solved in
an admittedly self-consistent manner.
In particular,
tidal stripping of less metal-enriched
ISM during dissipative galaxy merging  
is found to play a vital role in determining mean and radial 
chemical properties of ISM
in elliptical galaxies.
The present study accordingly implies that the origins of metal-poor gaseous halo and
negative metallicity gradients of ISM 
in  an elliptical galaxy can be  closely associated with gas-rich galaxy merging
at the epoch of elliptical galaxy formation.


\acknowledgments

We are grateful to the referee for valuable comments, which contribute to improve
the present paper.
K.B. thanks to the Japan Society for Promotion of Science (JSPS) 
Research Fellowships for Young Scientist.

\clearpage
\begin{deluxetable}{ccccc}
\footnotesize
\tablecaption{MODEL PARAMETERS \label{tbl-1}}
\tablewidth{0pt}
\tablehead{
\colhead{Model} & 
\colhead{$M_{\rm d}/M_{\odot}$} & 
\colhead{$R_{\rm chem}$} & 
\colhead{Comments}}
\startdata

Model 1 & $10^{10}$ & 0.1 & standard model  \\
Model 2 & $10^{12}$ & 0.1 & massive disks  \\
Model 3 & $10^{10}$ & 0.4 & larger mixing length \\
Model 4 & $10^{10}$ & 0.1 & stronger  supernova feedback \\
Model 5 & $10^{10}$ & 0.1 & multiple merger \\

\enddata

\end{deluxetable}


\figcaption{
Radial distribution of mean gaseous metallicity
$<Z>$ and $<Fe>$ (in units of solar metallicity
${<Z>}_{\odot}$ and ${<Fe>}_{\odot}$)
at $T=16T_{\rm dyn}$ (1.4 Gyr) 
in a remnant of a pair galaxy merger 
for Model 1.
Here, the distance ($R$) from the center of the merger remnant of Model 1 
is measured in units of initial disk size ($R_{\rm d}$).
Note that the merge remnant has steep  negative metallicity gradients
both for Z and Fe.
\label{fig-1}}

\figcaption{
The dependence of mean stellar metallicity ($<Z_{\ast}>$
and $<Fe_{\ast}>$) on mean gaseous one 
($<Z>$ and $<Fe>$) at $T=8, 12, 16 T_{\rm dyn}$
in the  merger remnant of Model 1 
for the region $R \leq R_{\rm eff}$ (crosses), 
$R \leq 2.5R_{\rm eff}$ (open triangles),
$R \leq 10.0R_{\rm eff}$ (open squares)
and the whole region (open  circles).
Here $R_{\rm eff}$ represents the effective radius of
the merger remnant in Model 1.
Since stellar and gaseous metallicity is 
in general larger for the results with larger $T$, 
the results located 
in the righter side represent those with larger $T$. 
For comparison, a boundary where $<Z>$ ($<Fe>$) is equal to
$<Z_{\ast}>$ ($<Fe_{\ast}>$) is given
by dotted lines.
This figure shows  time-evolution  of  the degree of difference in mean metallicity between
stellar components and gaseous ones (ISM). 
Note that $<Z>$ is larger than $<Z_{\ast}>$ for $R \leq R_{\rm eff}$
whereas $<Z>$ is discernibly smaller  than $<Z_{\ast}>$ for the whole region.
\label{fig-2}}

\figcaption{The same as Figure 1 but for 
a multiple merger model in Model 5.
\label{fig-3}}

\figcaption{The same as Figure 2 but for 
a multiple merger model in Model 5.
\label{fig-4}}

\figcaption{The  $<Z>-<Z_{\ast}>$ (upper)
and $<Fe>-<Fe_{\ast}>$  (lower) relationships of 
merger remnants at $T = 16.0 T_{\rm dyn}$
for the region $R \leq R_{\rm eff}$ (filled marks), 
and the whole region (open marks) in all models (Model 1 $\sim$ 5). 
Triangle, squares, pentagons, hexagons, and
circles represent
the results of Model 1, 
those of Model 2,  those of Model 3,
those of Model 4, and those of Model 5.
For comparison, a boundary where $<Z>$ ($<Fe>$) is equal to $<Z_{\ast}>$
($<Fe_{\ast}>$) is given
by dotted lines.
\label{fig-5}}

\end{document}